\documentclass[prl,twocolumn,superscriptaddress]{revtex4}

\usepackage{amsmath,amsfonts,amssymb,amstext,amsthm}
\usepackage{graphicx}
\usepackage[colorlinks,citecolor=blue]{hyperref}

\def\>{\rangle}
\def\<{\langle}
\def\({\left(}
\def\){\right)}

\begin{document}

\title{Adiabatic Cluster State Quantum Computing}

\author{Dave Bacon}
\affiliation{Department of Computer Science \& Engineering, University of Washington, Seattle, WA 98195}
\affiliation{Department of Physics, University of Washington, Seattle, WA 98195}
\author{Steven T. Flammia}
\affiliation{Perimeter Institute for Theoretical Physics, Waterloo, Ontario, N2L 2Y5 Canada} 
\date{\today}

\begin{abstract}
Models of quantum computation are important because they change the physical requirements for achieving universal quantum computation (QC).  For example, one-way QC requires the preparation of an entangled ``cluster'' state followed by adaptive measurement on this state, a set of requirements which is different from the standard quantum circuit model.  Here we introduce a model based on one-way QC but {\em without measurements\/} (except for the final readout), instead using adiabatic deformation of a Hamiltonian whose initial ground state is the cluster state.  This opens the possibility to use the copious results from one-way QC to build more feasible adiabatic schemes.
\end{abstract}

\maketitle

Computers that can exploit the laws of quantum theory can, in principle, outperform today's classical computers.  For example, quantum computers can efficiently factor~\cite{Shor:94a}, something classical computers are thought incapable of doing.  
Motivated by this fact, a vast amount of ongoing research focuses on figuring out exactly how to build a quantum computer.  In addition to different physical mediums for implementing QC, numerous different {\em models} for how to achieve QC have been proposed.  While to date each of these models provides the same computational power, they differ substantially on the requirements they put on the physical hardware.  The most widely used model of QC is the quantum circuit model, but other models include one-way (or measurement-based) QC~\cite{Raussendorf:01a}, holonomic QC~\cite{Zanardi:99d}, universal adiabatic QC~\cite{Aharonov:04a}, and topological QC~\cite{Kitaev:03a}.  Here we propose a new model of computing which combines ideas from all of these models.  In particular we demonstrate how one can perform one-way QC adiabatically.  

One-way QC~\cite{Raussendorf:01a} is a method for QC in which one creates a specific, fixed entangled state of a quantum many-body system and then computes via a series of local measurements on the subsystems.  The choice of measurements correspond to unitary gates enacted in the QC and these measurements are adaptive: that is, the exact measurement being executed depends on the previous measurement results.  One set of states which can be used for one-way QC are the class of so-called cluster states~\cite{Nielsen:06a}.  Cluster states are defined for any graph, though not all graphs allow for universal one-way QC.  A cluster state can be defined as a stabilizer code state.   Equivalently, there is a Hamiltonian with at most $(d+1)$-qubit interactions, where $d$ is the maximum degree of the graph, whose ground state is the cluster state (one can replace this Hamiltonian with another involving only $2$-qubit interactions while retaining the cluster state as an approximate ground state~\cite{Bartlett:06a}.)  Thus one could imagine engineering a physical system with this Hamiltonian, cooling the system to its ground state, and then doing measurements that enact the cluster state QC.  Here we show that instead of performing these measurements one can instead simply adiabatically turn on appropriate local fields while turning off portions of the cluster state in order to perform the QC.  Thus we can dispense with measurements in the one-way model (except, of course, for the final readout) and instead use adiabatic evolutions to enact one-way QC.  This model provides many of the advantages of adiabatic control; in particular it retains robustness to deformations of the specific adiabatic path traversed during the open-loop holonomic evolution~\cite{Kult:06a}.  

\paragraph{Adiabatic dragging.}

The main tools we use in this paper are adiabatic changes in a Hamiltonian.  Suppose initially we have a system with Hamiltonian $H_i$ and the system is in an energy eigenstate.  Then we evolve the system under a time-varying Hamiltonian over a time period $0 \leq t \leq T$ as $H(t)=f\left(s\right) H_i +g \left(s \right) H_f$ where $f(0)=g(1)=1$, $f(1)=g(0)=0$ and $s={t \over T}$ is a scaled time.  If we vary this evolution smoothly and there are no level crossings, then it is always possible to choose a $T$ large enough such that at the end of this evolution we will be in the eigenstate of $H_f$ which is continuously connected to the initially prepared eigenstate.  In particular, if we choose $T$ to be on the order of the minimum energy gap between the instantaneous eigenstate of $H(t)$ and the nearest eigenstates, then with high probability at the end of the above evolution the system will be in the connected eigenstate of the final Hamiltonian~\cite{Schaller:06a}.  We will call such a setup and evolution an {\em adiabatic dragging}.  Recently, adiabatic dragging between Hamiltonians with energy eigenstates that are degenerate and are quantum error-correcting codeword states has emerged as a powerful primitive for building a quantum computer~\cite{Oreshkov:09a,Oreshkov:09b,Bacon:09a}.  Here we extend these ideas to one-way QC.

\paragraph{1D degenerate cluster-state model.}

Begin by considering a line of $n$ qubits and a degenerate variation on the one-dimensional cluster state.  In particular define the following $n-1$ commuting operators
\begin{equation}
S_i=[Z]_{i} [X]_{i+1} [Z]_{i+2}, 1 \leq i \leq n-2,  S_{n-1}=[Z]_{n-1} [X]_n,  \nonumber
\end{equation}
where $X$ and $Z$ are the corresponding Pauli operators and we use the notation $[P]_i$ to denote the operator $P$ acting on the $i$th physical qubit.  These are $n-1$ out of the $n$ operators usually used to define a cluster state~\cite{Nielsen:06a}.  Define now the stabilizer code corresponding to these operators as the common $+1$ eigenstates of all of the $S_i$, i.e. $|\psi\rangle$ such that $S_i |\psi\rangle=|\psi\rangle$.  By standard results in the theory of stabilizer codes~\cite{Gottesman:97a}, this code space is two dimensional (encodes a qubit.)  We can define the logical operators for this encoded qubit as
\begin{equation}
\bar{X}=[X]_1 [Z]_2 \quad {\rm and} \quad  \bar{Z}=[Z]_1. \label{eq:pauli}
\end{equation}
Now consider the Hamiltonian
\begin{equation}
H_0=-\Delta \sum_{i=1}^{n-1} S_i. \label{eq:initialham}
\end{equation}
Since the $S_i$ all commute and have eigenvalues $\pm 1$, the ground state subspace of this Hamiltonian is the $+1$ common eigenstate of the $S_i$'s or, in other words, the encoded qubit defined above.  Note that quantum information in the degenerate ground state can be accessed by measuring or manipulating the encoded Pauli operators which are themselves localized on the first two qubits.

Now suppose that we adiabatically turn on a local field along the $-[X]_1$ direction while turning off the $S_1$ term in $H_0$, which anticommutes with $[X]_1$.  In particular consider adiabatic dragging from $H_0$ to $H_0+\Delta (S_1 - [X]_1)$.  Notice that while $\bar{X}$ commutes with $[X]_1$, $\bar{Z}$ does not commute with $[X]_1$.  However because we are in the $+1$ eigenspace of each $S_i$, instead of defining the logical $\bar{Z}$ as we have done above in Eq.~(\ref{eq:pauli}) we could also define the encoded $Z$ as $\bar{Z}^\prime=\bar{Z} S_1= [X]_2 [Z]_3$.  If we do this, then the encoded qubit commutes with the terms we are turning on and off ($S_1$ and $[X]_1$.)  Thus the quantum information in this encoded subspace is not touched.  However since $S_1$ anticommutes with $[X]_1$, the information in $S_1$ is changed.  To see how this evolution proceeds, we can consider a code in which we promote $S_1$ into an encoded Pauli $Z$ operator and $[X]_1$ is its conjugate encoded $X$ operator.  The adiabatic evolution is then simply between these the two encoded Pauli operators (i.e. from an encoded $-\Delta \bar{Z}_a$ to an encoded $-\Delta \bar{X}_a$ where $a$ denotes this newly defined encoded qubit.)  Such an evolution has no level crossing and an energy gap for reasonable adiabatic interpolations which is proportional to $\Delta$.   Thus at the end of this evolution we will be in the $+1$ eigenstate of $[X]_1$ along with all the remaining $S_i$.  In other word we are in the stabilizer code with stabilizer generators $[X]_1,S_2,S_3,\dots, S_{n-1}$.  The information in the degenerate subspace, which originally was represented via the encoded operators $\bar{X}=[X]_1[Z]_2$ and $\bar{Z}=[Z]_1$ is now represented by $\bar{X}^\prime=[X]_1 [Z]_2$ and $\bar{Z}^\prime=[X]_2 [Z]_3$.  However, since we are in the $+1$ eigenstate of $[X]_1$ this is equivalent to the encoded operator $\bar{X}^{\prime \prime}= [Z]_2$ and $\bar{Z}^{\prime \prime}=[X]_2 [Z]_2$.  In other words the information which was originally encoded in the first two qubits, after the above adiabatic dragging, will be in the second and third qubit.  Using the same logical Pauli encoding (logical $X$ is $[X]_i [Z]_{i+1}$ and logical $Z$ is $[Z]_i$) we see that a Hadamard gate has been applied to this information.  Thus, by turning on a $[X]_1$ term on the first qubit while turning off the term in the Hamiltonian with which it anti-commuted, we have effectively moved this information one step down the line, and applied a Hadamard gate to the quantum information.

Proceeding inductively, if we first adiabatically turn on $[X]_1$, then $[X]_2$, etc, while turning off the corresponding anticommuting term in the original Hamiltonian we will end up with the qubit which was originally localized to one end of the line moved to the other end of the line, along with a sequence of Hamadard gates applied to this qubit. Throughout this piecewise evolution the energy gap will remain constant because each successive adiabatic dragging acts independently.  If we proceed to turn on each $[X]_i$ all the way up to the $(n-1)$st qubit, the information originally encoded into the first two qubits will end up exactly on the last qubit.  In other words after this evolution, $\bar{X}$ is mapped to $[X]_n$ and $\bar{Z}$ is mapped to $[Z]_n$ if the chain is odd length and $\bar{X}$ is mapped to $[Z]_n$ and $\bar{Z}$ is mapped to $[X]_n$ otherwise --- these differences arising from whether an even or odd number of Hadamards have been applied to the encoded qubit.

\paragraph{Single qubit gates.}

We now show how to modify the above setup such that in addition to propagating a single qubit of information down the one dimensional system, we also apply gates other than the Hadamard gate to the qubit.  This scheme is motivated directly by the one-way QC model where instead of measuring the qubit along the $X$ direction to propagate the information, we measure along a rotated direction, $M(\theta)=\cos(\theta)X + \sin(\theta)Y$.  Importantly, however, our scheme proceeds {\em without adaptive operations}.  Consider mimicking the above scheme, but instead of turning on successive $-\Delta [X]_i$s while turning off the appropriate anticommuting terms in $H_0$ (the $-\Delta [Z]_i [X]_{i+1} [Z]_{i+2}$ terms) we instead turn on successive $-\Delta M_i$ terms where $M_i=[M(-\theta_i)]_i$ is a set of rotated local fields, $1 \leq i \leq n-1$.  We claim that this will take the qubit localized to one end of the line and propagate it to the other end of the line while applying a gate dependent on the choice of $\theta_i$. 

To analyze this scheme it is easier to work in a frame of reference in which the $i$th qubit has been rotated by $U(\theta_i) = \exp(-i \theta_i [Z]_i/2)$.  It is convenient to take $\theta_1=0$, which we will now assume.  Consider again $n$ qubits on a line and define now the {\em rotated} stabilizer code operators:
\begin{eqnarray}
T_i&=&[Z]_{i} [X^{U_{i+1}}]_{i+1} [Z]_{i+2}, \quad 1 \leq i \leq n-2, \nonumber \\ T_{n-1}&=&[Z]_{n-1} [X]_n \,,
\end{eqnarray}
where we use the superscript to denote conjugation, $P^U=UPU^\dagger$, and $U_i=U(\theta_i)$.  Note that this conjugation does not change the fact that these operators commute and square to identity, and therefore we can again define a codespace as the joint $+1$ eigenspace of these operators.  Let $H_0$ be the initial Hamiltonian for our system as in Eq.~(\ref{eq:initialham}), but now with the rotated stabilizer operators $T_i$ substituted for $S_i$.  Again, initially we can define the information in the degenerate subspace as localized to the first two qubits with $\bar{X}=[X]_1 [Z]_2$ and $\bar{Z}=[Z]_1$.  Now imagine adiabatically dragging $H_0$ to $H_0+\Delta (T_1 - [X]_1)$, then dragging to $H_0+\Delta(T_1+T_2-[X]_1-[X]_2)$, etc.  We claim that at the end of this scheme we will end up with the quantum information in $\bar{X}$ and $\bar{Z}$ propagated to the last qubit with a gate dependent on $\theta_i$ applied to this information.  

To see this, we proceed in three steps.  First we will show that using the rotated stabilizer operators it is possible to write the logical qubit in a form where each $X_i$ (except $i=n$) commutes with this information.  Define the following operators for $\alpha,\beta \in \{X,Y,Z\}$: 
$\bar{\alpha}_i = \sum_\beta (P^{\alpha,\beta}_i [\beta]_i)^{C_{i,i+1}} $
where $C_{i,i+1}$ is the controlled phase gate between the $i$ and $(i+1)$st qubits except when $i=n$ in which case we define $C_{n,n+1}=I$.  We claim that these new Pauli operators are, under the rotated stabilizer code generated by the $T_i$'s, equivalent to the logical operators $\bar{X}=[X]_1 [Z]_2$,  $\bar{Y}=[Y]_1 [Z]_2$, and $ \bar{Z}=[Z]_1$, with the condition that the $P^{\alpha,\beta}$'s are a sum of products of $[X]_j$ operators for $j<i$.  This can be proven inductively.  The base case corresponds to $P^{\alpha,\beta}=\delta_{\alpha,\beta}I$ where $\bar{X}_1=\bar{X}$ and $\bar{Z}_1=\bar{Z}$.  Now assume the hypothesis is true for the $i$th operators.  Examine, for example, $\bar{X}_i$ and expand the controlled-phase:
 \begin{equation}
\bar{X}_i=P^{X,X}_i [X]_i [Z]_{i+1} +P^{X,Y}_i [Y]_i [Z]_{i+1}+P^{X,Z}_i [Z]_i.
\end{equation}
Recall that the $T_i$ operators act as identity on the codespace and thus can be inserted into this sum in any manner to yield any equivalent operator (over the code.)  Left multiplying $\bar{X}_i$ by $T_i$ for the last two terms yields
\begin{eqnarray}
\bar{X}_i&=&P^{X,X}_i [X]_i [Z]_{i+1} +P^{X,Z}_i [X^{U_{i+1}}]_{i+1} [Z]_{i+2} \nonumber \\
&&-i P^{X,Y}_i [X]_i [X^{U_{i+1}}]_{i+1}  [Z]_{i+1} [Z]_{i+2} 
\end{eqnarray}
Expanding out $X^{U_{i+1}}$, we find that
\begin{eqnarray}
P^{X,X}_{i+1}&=&  \cos(\theta_{i+1})  P^{X,Z}_i +   \sin (\theta_{i+1}) [X]_i P^{X,Y}_i\nonumber \\
P^{X,Y}_{i+1}&=&  \sin(\theta_{i+1}) P^{X,Z}_i - \cos(\theta_{i+1}) [X]_i P^{X,Y}_i \nonumber \\
P^{X,Z}_{i+1} &=& [X]_i  P^{X,X}_i 
\end{eqnarray}
Similar relations hold for $\bar{Y}_{i+1}$ and $\bar{Z}_{i+1}$ with the important property that the new $P^{\alpha,\beta}_{i+1}$s are functions of the previous $P^{\alpha,\beta}_{i}$s and $[X]_i$s.  This proves our statement.

But these expressions also prove much more.  In particular if we restrict the above equivalence to the $+1$ subspace of $[X]_i$, then we see (when we calculate out all nine new $P^{\alpha,\beta}_{i+1}$'s) that the relationship between the $\bar{\alpha}_i$ and $\bar{\alpha}_{i+1}$ is $\bar{\alpha}_{i+1}=\bar{\alpha}_{i}^{U_{i+1} H}$.  In other words, with this restriction, the effect on the encoded quantum information in this new form is as if the gate $U_{i+1}H$ has been applied to the quantum information.  Further note that in the procedure we have described for adiabatically dragging the initial Hamiltonian, we are always turning off a $-\Delta T_i$ while turning on a $-\Delta [X]_i$.  Then not only does $[\alpha]_{i+1}$ commute with these terms (because the $P^{\alpha,\beta}_{i+1}$ is made up entirely of a product of $[X]_j$'s with $j<i+1$), and thus is untouched by the evolution, but by an argument identical to the untwisted Hamiltonian case, we end each such dragging in the $+1$ eigenvalue of $-\Delta [X]_i$.  Thus we end up exactly in the subspace where the gate $U_{i+1}H$ has been applied and the quantum information shifted one site down the chain for each such adiabatic dragging.  The final effect for the turning on all $n-1$ $[X]_i$ in order is that the sequence of gates $H \prod_{i=n-2}^{2} (U_{i+1} H) $ is applied to the quantum information.
  
In recap, we have shown that by starting with a Hamiltonian which is a negative sum of twisted stabilizer operators $T_i$ and then turning off the $T_i$'s while turning on the $[X]_i$'s sequentially, we have enacted a gate which depends on the angles $\theta_i$.  This is equivalent to using the standard cluster state Hamiltonian from Eq.~(\ref{eq:initialham}) with the unrotated $S_i$ stabilizer operators as the initial Hamiltonian and using rotated magentic fields $[M(-\theta_i)]_i$ for the piecewise final Hamiltonians.  Note that we did not work in a rotating frame for the final qubit and therefore the information ends up exactly on the last qubit of this evolution.  Throughout this piecewise evolution the energy gap is constant (independent of the length of the chain.)   The gates enacted are universal for single qubit gates.

\paragraph{State preparation.}

In the previous section we enacted gates on the degenerate ground state of a Hamiltonian.  We now show how it is possible to prepare quantum information in a particular state, with the Hamiltonian non-degenerate, and then propagate the information down the line while turning the Hamiltonian into one with a degenerate ground state where this encoded information lives.  Consider, for example, our original Hamiltonian in Eq.~(\ref{eq:initialham}) but now with the full cluster state Hamiltonian $H_0^\prime=H_0-\Delta S_0$ where $S_0=[X]_1 [Z]_2$.  The ground state of $H_0^\prime$ is now not degenerate and corresponds, in our previous picture of $H_0$ to being in the $+1$ eigenstate of $\bar{X}$.  Consider first adiabatically dragging $H_0^\prime$ to $H_0^\prime+\Delta(S_1-[X]_1)$.  At the end of this evolution we will be in the $+1$ eigenspace of $[X]_1$ as before. Since we started in the $+1$ eigenspace of $\bar{X}$ we will be in the $+1$ eigenspace of $\bar{X}^\prime=[Z]_2$.  Next adiabatically drag the Hamiltonian to $H_0^\prime+\Delta(S_0+S_1+S_2-[X]_1-[X]_2)$.  Notice that we have to turn off two stabilizer generators while turning on a single field.  This implies that we must {\em increase the degeneracy\/} of the ground state.  We will see that this second dragging, despite increasing the degeneracy, ends with the system in the $+1$ eigenstate of the $\bar{X}^{\prime \prime}=[X]_3 [Z]_4$.

To see this note that while both $S_1$ and $S_2$ do not commute with $[X]_2$, $S_1 S_2$ does.  Thus the eigenvalue of $S_1S_2$ is preserved while turning on $[X]_2$.  If we then rewrite $S_1+S_2$ as $S_1 (I+S_1 S_2)$, then if we are in the $-1$ eigenspace of $S_1S_2$ then this term vanishes, but if we are in the $+1$ eigenspace then in this space the operator effectively acts as $2S_1$ (or equivalently $2S_2$).  We can then consider the code where we promote $S_1$ to an encoded $Z$ operator and $[X]_2$ to an encoded $X$ operator, and then at the end of the evolution we will be in the $+1$ eigenstate of $[X]_2$, and we are also in the $+1$ eigenstate of $S_1 S_2$ (due to this operator commuting with $[X]_2$.)  Translating this into the coding language, we are in the $+1$ eigenstate of a stabilizer code with generators $[X]_1,[X]_2,[X]_3 [Z]_4,S_3,\dots,S_{n}$, which is equivalent to saying that we are in the $+1$ eigenstate of the 1D cluster state with $n-2$ qubits but prepared in the $+1$ eigenstate of the encoded $\bar{X}$ at one end of this chain.  If we wish to apply gates to this information, we can proceed as above by applying rotated local fields or rotating the stabilizer Hamiltonian.  It is important to realize that the above evolution has gone from a non-degenerate to a degenerate ground state, so that the energy gap vanishes.  However over the subspaces defined by the conserved quantity $S_1 S_2$ the energy gap is constant and thus the adiabatic theorem holds.  In fact, the same situation occurs in the creation of anyons in topological QC~\cite{Bacon:09b}.


\paragraph{Two-qubit gates.}

Let us now show how to apply two-qubit gates.  The idea, just as in one-way QC, is to use a Hamiltonian which has a coupling between two chains which support single qubits.  To see how this works let us analyze a cluster state Hamiltonian with a degenerate ground subspace and a single coupling between two encoded qubits.  Consider the six-qubit initial Hamiltonian  
\begin{equation}
H_2=-\Delta( [Z]_{1,a} [X]_{2,a} [Z]_{3,a} [Z]_{2,b} + [Z]_{2,a} [X]_{3,a}) +(a \leftrightarrow b) \nonumber
\end{equation}
where the encoded qubits will be associated with $a$ and $b$ and $(a \leftrightarrow b)$ denotes same term with the $a$ and $b$ labels reversed.  This Hamiltonian is degenerate, but now there are two qubits of degeneracy, corresponding to logical operators $\bar{X}_\gamma=[X]_{1,\gamma} [Z]_{2,\gamma}$ and $\bar{Z}_\gamma=[Z]_{1,\gamma}$ with $\gamma  \in \{a,b\}$.  Now suppose that we turn on $-\Delta ([X]_{1,a}+[X]_{1,b}+[X]_{2,a}+[X]_{2,b})$ while turning off  $H_2$ (we could proceed by turning each of these on separately and achieve similar results.)  Using the four stabilizer terms in the Hamiltonian above we can rewrite the encoded operators as $\bar{X}^\prime_\gamma=[X]_{1,\gamma} [X]_{3,\gamma}$ and $\bar{Z}^\prime_\gamma=[X]_{2,\gamma} [Z]_{3,\gamma} [X]_{3, \neg \gamma}$ where $\neg a=b$ and $\neg b =a$.  Using an argument similar to the single-qubit gates, we will end up in the $+1$ eigenstate of the $X_{i,\gamma}$ operators, $i \in \{1,2\}$.  Over this eigenspace, the logical operators become $\bar{X}_f=[X]_{3,\gamma}$ and $\bar{Z}_f = [Z]_{3,\gamma} [X]_{3,\bar{\gamma}}$.  This is equivalent to performing a Hadamard on each encoded qubit, a controlled phase gate between these qubits, and then a Hadamard on each qubit again.

\paragraph{Adiabatic Cluster State QC (ACSQC).}

We now see how to build a quantum computer using piecewise adiabatic evolutions from a Hamiltonian whose ground state is a cluster state to a Hamiltonian consisting of local fields (we note that this initial state can also be piecewise adiabatically prepared~\cite{Schaller:08a}.)  Consider a quantum circuit made up of gates from a universal gate set such as $\{H U({\pi \over 4}),H,(H\otimes H) C_{i,j}\}$ (other sets are also possible) along with the preparation in the $+1$ eigenstate of the Pauli $X$ operator.  Then one can map the graph of this circuit onto a cluster state graph using the above elements in such a way that one can also prescribe local fields which, when turned on piecewise, enact the quantum circuit (or equivalently one can us a twisted cluster state Hamiltonian and local fields all along $X$.)  

\paragraph{Conclusion.}

We have shown how to perform one-way QC on a cluster state using only piecewise adiabatic evolutions.  This scheme shares many of the traits of the recently introduced primitive of adiabatic gate teleportation \cite{Bacon:09a}: it has a robustness to the adiabatic path, for example.  Further, as in~\cite{Bacon:09a} we can use perturbation theory gadgets~\cite{Bartlett:06a} to implement this entire scheme using only two-qubit interactions instead of the four-qubit interactions we have presented; it would be interesting to make this calculation explicit.  Our model shows the novelty of starting with a global entangled ground state and then piecewise turning on local fields to do QC.  We have also shown how it is possible to use cluster states and their parent Hamiltonians to perform QC without resorting to adaptive measurements.  ACSQC thus opens up a new way to adapt the numerous results of one-way QC to viable adiabatic architectures.

While preparing this manuscript we learned of similar results for single qubit circuits using the AKLT state~\cite{Renes:09a}.

\acknowledgments

DB was supported by the NSF under grants 0803478, 0829937, and 0916400 and by DARPA under QuEST grant FA-9550-09-1-0044.  STF and research at Perimeter are supported by the Government of Canada through Industry Canada and by the Province of Ontario through the Ministry of Research~\& Innovation.

\vspace{-10pt}

\end{document}